\RequirePackage{ifpdf}
\ifpdf 
\documentclass[pdftex]{sigma}
\else
\documentclass{sigma}
\fi

\begin{document}

\allowdisplaybreaks

\renewcommand{\thefootnote}{$\star$}

\renewcommand{\PaperNumber}{018}

\FirstPageHeading

\ShortArticleName{The Integrability of New Two-Component KdV Equation}

\ArticleName{The Integrability of New Two-Component\\  KdV
Equation\footnote{This paper is a contribution to the Proceedings of
the XVIIIth International Colloquium on Integrable Systems and Quantum
Symmetries (June 18--20, 2009, Prague, Czech Republic).  The full
collection is
available at
\href{http://www.emis.de/journals/SIGMA/ISQS2009.html}{http://www.emis.de/journals/SIGMA/ISQS2009.html}}}

\Author{Ziemowit POPOWICZ}

\AuthorNameForHeading{Z.~Popowicz}

\Address{Institute for Theoretical Physics, University of Wroclaw,
Wroclaw 50204, Poland}
\Email{\href{mailto:ziemek@ift.uni.wroc.pl}{ziemek@ift.uni.wroc.pl}}
\URLaddress{\url{http://www.ift.uni.wroc.pl/~ziemek/}}

\ArticleDates{Received October 19, 2009, in f\/inal form February 04,
2010;  Published online February 12, 2010}

\Abstract{We consider  the bi-Hamiltonian  representation of the
two-component coup\-led KdV equations discovered by
Drinfel'd and Sokolov and rediscovered by Sakovich and Foursov.
Connection of this equation with the supersymmetric
Kadomtsev--Petviashvilli--Radul--Manin hierarchy is presented. For
this new supersymmetric equation the Lax representation and odd
Hamiltonian structure is given.}


\Classification{35J05; 81Q60}

\section{Introduction}

The scalar KdV equation admits various generalizations to the
multif\/ield case and  have been often considered in the literature
\cite{swinia,gora,anton,mau,cztery}. However, the present
classif\/ication of such systems is not complete and depends on the
assumption which we made on the very beginning.

Svinolupov \cite{swinia} has introduced the class of equations
\begin{gather}\label{eq1}
 u^{i}_{t} = u^{i}_{xxx} + a^{i}_{j,k} u^j u_x^{k},
\end{gather}
where $i,j,k=1,2,\dots ,N $, $u_{i}$ are functions depending on the
variables $x$ and $t$ and $a^{i}_{j,k}$
are constants.  The $a^{i}_{j,k}$    satisfy the same relations as the
structural constants of Jordan algebra
\begin{gather*}
 a^{n}_{j,k}\big(a^{j}_{n,r}a^{r}_{m,s} - a^{i}_{m,r}a^{r}_{n,s}\big)
+ {\rm cyclic}(j,k,m)=0.
\end{gather*}
These  equations possess inf\/initely many higher generalized symmetries.

G\"{u}rses and Karasu \cite{gora} extended the Svinolupov
construction   to the  system of equations
\begin{gather}\label{eq3}
 u^{j}_{t} = b^{i}_{j}u^{i}_{xxx} + s^{i}_{j,k} u^j u^{k}_{x},
\end{gather}
where $b^{i}_j$, $s^{i}_{j,k}$ are constants.

In general, existence of inf\/initely many conserved quantities is
admitted as the def\/inition of integrability. This implies existence of
inf\/initely many generalized
symmetries. G\"{u}rses and Karasu, in order to check the integrability
of the system of equations \eqref{eq3},
assumed that the system is integrable if it admits a recursion operator.
Assuming the general form  of the second and fourth order recursion operator
they found the conditions on the coef\/f\/icients $b^{i}_{j}$, $s^{i}_{j,k}$
so that the equations in \eqref{eq3} are integrable.

A quite dif\/ferent generalization of multicomponent KdV system has been found by
Anto\-no\-wicz and Fordy \cite{anton}  considering the energy  dependent
Schr\"{o}dinger operator. Ma~\cite{mau} also presented a
multicomponent KdV system considering decomposable hereditary
operators.

Several years ago Foursov \cite{cztery} found the conditions on the
coef\/f\/icients $b^{i}_{j}$,
$s^{i}_{j,k}$ under which the two-component system \eqref{eq3}
possesses  at least 5
generalized symmetries  and  conserved quantities.
He carried out computer algebra computations and found  that there are
f\/ive such systems which are  not symmetrical and not triangular. Three
of them are known to be integrable, while two of them are new.
Foursov conjectured that these two new systems should be integrable.
However, it appeared that one of these new  systems is not new and has
been known for many years. Drinfel'd and Sokolov in 1981
\cite{sokal}  presented the Lax pair  for one of these new equation
and hence this equation is integrable.

In this paper we present the bi-Hamiltonian formulation and recursion
operator for  the \textit{new} equation.
These results have been obtained during the study of the so called
supersymmetric Manin--Radul hierarchy.
The application of the supersymmetry to the construction of new
integrable systems appeared almost in parallel to
the use of this symmetry in the quantum f\/ield theory. The quantum
f\/ield theories with exact correspondences  between bosonic and
fermionic helicity states are not the only basic ingredients for
superstring theories, but have been
utilised both in theoretical and experimental research in particle physics.
The f\/irst results, concerned the construction of classical f\/ield
theories with  fermionic and bosonic f\/ields depending on time and one
space variable, can be found in~\cite{Kuper0,Gurses,Kulish,Manin}. In
many cases, addition of fermion f\/ields does not guarantee that the
f\/inal theory becomes supersymmetric invariant. Therefore this method
was named as a fermionic extension in order to distinguish it from the
fully supersymmetrical method which was developed
later~\cite{Mat1,Mat2,Mat3,Masud1,Pop1}. There are many recipes how
some classical  models could be embedded  in  fully supersymmetric
superspace. The main idea is simple: in order to get such
generalization we should construct a supermultiplet containing  the
classical functions. It means that we have to add to a system of $k$
bosonic equations $kN$ fermions and $k(N-1)$ bosons ($k=1,2,\dots$, $
N=1,2,\dots$) in such  a way that they create superf\/ields. Now working
with this supermultiplet we can step by step  apply  integrable
Hamiltonians methods to our considerations depending on what we would
like to construct.

Manin and Radul in 1985 \cite{Manin},  introduced a new system of
equations for an inf\/inite set of even and odd functions, depending on
an even-odd pair of space variables and even-odd times. This system of
equations
now called the Manin--Radul supersymmetric Kadomtsev--Petviashvili
hierarchy (MR-SKP). It appeared that  this
hierarchy contains the supersymmetric generalization of the
Korteweg--de Vries equation, the Sawada--Kotera equation and as we
show in this paper two-component coupled KdV equations discovered by
Drinfel'd--Sokolov.

\section{Two-component KdV systems}\label{section2}

Let us consider a system of two equations
\begin{gather*}
u_t  =  F[u,v],  \qquad
v_t  =  G[u,v],
\end{gather*}
where $F[u,v] =F(u,v,u_x,v_x,\dots)$ denotes a dif\/ferential polynomial
function of $u$ and $v$.

By the triangular system we understand such system which involves
either an equation depending only on $u$ or an equation depending only
on $v$ while by the symmetrical we understand such system in which
$G[u,v]= F[v,u]$.

\begin{definition}
A system of $t$-independent evolution equations
\begin{gather*}
u_t  =  Q_{1}[u,v],  \qquad
v_t  =  Q_{2}[u,v]
\end{gather*}
is said to be a generalized symmetry of  \eqref{eq1}  if their f\/lows
formally commute
\begin{gather*}
{\bf D}_{\bf K}({\bf Q}) - {\bf D}_{{\bf Q}}({\bf K}) = {\bf 0}.
\end{gather*}
Here ${\bf Q}=(Q_1,Q_2)$, ${\bf K}[u,v] = (F[u,v],G[u,v])$, and ${\bf
D}_{\bf K}$ denotes the Fr\'{e}chet derivative.
\end{definition}

The f\/irst three  systems in the Foursov classif\/ication are known to be
integrable equations and are
\begin{gather*}
 u_t =  u_{xxx} + 6uu_{x} -12 vv_x,  \\
 v_t =  -2v_{xxx} - 6uv_x;
\\
  u_t  =   u_{xxx} + 3uu_{x} + 3vv_x,   \\
 v_t  =  u_xv + uv_x;
\\
  u_t  =   u_{xxx} + 2vu_{x} + uv_x,  \\
 v_t  =  uu_x. 
\end{gather*}
The f\/irst pair of equations  is the Hirota--Satsuma system
\cite{hirek}, second is the Ito system \cite{itek},  third is the
rescaled Drinfel'd--Sokolov equation \cite{cztery}.

The fourth system of equations is a new one founded by Foursov
\begin{gather*}
  u_t  =   u_{xxx} + v_{xxx} + 2vu_{x} + 2uv_x, \nonumber  \\
 v_t  =  v_{xxx} - 9u u_x + 6 vu_x + 3uv_x + 2vv_x.
\end{gather*}
Foursov showed that this system  possesses generalized symmetries of
weights 7, 9, 11, 13, 15, 17  and~19,   as well as conserved densities
of weights 2, 4, 6, 8, 10, 12 and~14, and  conjectured that this
system is integrable and should possess inf\/initely many generalized
symmetries.

The last system in this classif\/ication is
\begin{gather}
  u_t  =   4u_{xxx} +3v_{xxx} + 4uu_x + vu_x + 2uv_x, \nonumber \\
 v_t  =  3u_{xxx} + v_{xxx} -4vu_x - 2uv_x - 2vv_x, \label{ds}
\end{gather}
and has been f\/irst considered many years ago by Drinfel'd and Sokolov
\cite{sokal} and
rediscovered by S.Yu.~Sakovich~\cite{sak1}.

Let us notice that the integrable Hirota--Satsuma  equation has the
following Lax representation \cite{ford}
\begin{gather*}
L  =  \big(\partial^{2} + u+v\big)\big(\partial^{2} + u-v\big), \qquad
\frac{\partial L}{\partial t}  = 4 \big[ L^{3/4}_{+} , L \big], 
\end{gather*}
while the integrable Drinfel'd--Sokolov  equation possesses the
following Lax representation \cite{gora,pop}
\begin{gather*}
L = \big(\partial^{3} + (u-v)\partial
+(u_{x}-v_{x})/2\big)\big(\partial^{3}
+(u+v)\partial+(u_x+v_x)/2\big), \qquad
\frac{\partial L}{\partial t}  =  4 \big[ L^{3/4}_{+} , L \big].
\end{gather*}

On the other side, the Lax operators of the Hirota--Satsuma equation
and of the Drinfel'd--Sokolov equation could be considered as special
reduced Lax operators of the fourth and sixth order respectively.
Indeed,  the Hirota--Satsuma  Lax operator could be rewritten  as
\begin{gather}\label{eq14}
 L=\partial^{4} + g_{2}\partial^2 + g_{1}\partial + g_{0},
\end{gather}
where
\begin{gather*}
 g_2 = 2u , \qquad   g_1=2(u_x - v_x) , \qquad    g_0=u_{xx} + u^2
-v_{xx} - v^{2}.
\end{gather*}

In this context, one can ask what kind of the equations follows from
the f\/ifth-order Lax operator which is
parametrised by  two functions of  same weight. Let us therefore
consider the following Lax operator
\begin{gather*}
 L=\partial^5 + h_2\partial^3 + h_3\partial^2+ h_4\partial + h_5,
\end{gather*}
where $h_i$, $i=2,3,4,5$ are polynomials in  $u$ and $v$ and their
derivatives of the dimension $i$.
Computing the Lax representation for this operator
\begin{gather*}
 \frac{\partial L}{\partial t} = 5 \big[L,L^{3/5}_{+}\big]
\end{gather*}
we obtained then
\begin{gather}\label{eq18}
 L = \big ( \partial^3 + \tfrac{2}{3}u\partial + \tfrac{1}{3}u_{x}\big
) \big (\partial^2 - \tfrac{1}{3} v \big )
\end{gather}
produces the system of equation  \eqref{ds}.

Let us notice that the Lax operator \eqref{eq14} is  factorized as the
product of two Lax operators.
The f\/irst one  is the Lax operator of the Kaup--Kupershmidt
equation while the second  is the Lax operator of the Korteweg--de
Vries equation. It is exactly the same Lax operator which has been
found by Drinfel'd and Sokolov \cite{sokal}.

Hence  we encounter the situation in which the  Lax operator of the
Korteweg--de Vries and
the Kaup--Kupershmidt equations can be used for construction of
additional equations.
This could be schematically presented as:

\smallskip

\centerline{\begin{tabular}{|c|c|c|}
\hline
\tsep{4pt}
 &$\tilde L_{\rm KdV}$&$\tilde L_{\rm KK}$ \bsep{2pt}\\
\hline
\tsep{4pt}
$L_{\rm KdV}$&Hirota--Satsuma& equations \eqref{ds} \bsep{2pt} \\
\hline
\tsep{4pt}
$L_{\rm KK}$&equations \eqref{ds} &Drinfel'd--Sokolov \bsep{2pt}\\
\hline
\end{tabular}}

\smallskip

\noindent
where $L_{\rm KdV}$, $\tilde L_{\rm Kdv} $ are two dif\/ferent Lax
operators of the Korteweg--de Vries equation while $L_{\rm KK} $ and
$\tilde L_{\rm KK}$ are two dif\/ferent
Lax operators of the Kaup--Kupershmidt equation.

\section{The recursion operator and bi-Hamiltonian structure}

From the knowledge of the Lax operator for evolution equations one can
infer a lot of properties
of these equations. The generalized symmetries are obtained by
computing the higher f\/low of the
Lax representation while the conserved charges follow from the trace
formula \cite{blacha}
of the Lax operator.

Using this technique we found f\/irst three conserved quantities  for
the equation \eqref{ds}
\begin{gather*}
H_1  =  \int dx \big( v^2 + 4u^2 + 6uv \big), \\ \nonumber
H_2  =  \int dx \big(495u_{4x}u-510u_{x}^2u+32u^4 + 2v^4
+630v_{4x}u + 180v_{4x}v-210v_{xx}vu - 210v_{x}^2u \\ \nonumber
 \phantom{H_2  =}{}  +75v_{x}^2v-525v_{x}u_{x}u
+14v^3u+28v^2u^2-105vu_{xx}u+56vu^3 \big ), \\ \nonumber
H_3  =  \int dx \big(182250u_{8x}u+769500u_{4x}u_{xx}u+445500u_{xxx}^2u+259200u_{xx}^2u^2\!+
223425u_{xx}u_{x}^2u   \\ \nonumber
 \phantom{H_3  =}{}
-104400u_{x}^2u^3+1344u^6+222750v_{8x}u+70875v_{8x}v-148500v_{6x}vu-160875v_{5x}u_{x}u
 \\ \nonumber
 \phantom{H_3  =}{}  -
594000v_{5x}v_{x}u-1113750v_{5x}v_{xx}u-128250v_{5x}v_{xx}v+54450v_{5x}v^2u-825v_{5x}vu^2
 \\ \nonumber
\phantom{H_3  =}{}
-742500v_{xxx}^2u\!-74250v_{xxx}^2v\!-61875v_{xxx}u_{xxx}u\!-217800v_{xxx}u_{x}u^2\!+267300v_{xxx}v_{x}vu
 \\ \nonumber
\phantom{H_3  =}{} +
70125v_{xx}^2u^2+19575v_{xx}^2v^2+163350v_{xx}^22vu-193050v_{xx}u_{x}^2u+297000v_{xx}v_{x}^2u
\\ \nonumber
\phantom{H_3  =}{}  +
17550v_{xx}v_{x}^2v+199650v_{xx}v_{x}u_{x}u-9900v_{xx}v^3u+15400v_{xx}vu^3-32175v_{x}^2u_{xx}u
\\ \nonumber
\phantom{H_3  =}{}  -
4400v_{x}^2u^3+3600v_{x}^2v^3-19800v_{x}^2v^2u+13200v_{x}^2vu^2-185625v_{x}u_{5x}u
\\ \nonumber
\phantom{H_3  =}{} +
188100v_{x}u_{xx}u_{x}u-79200v_{x}u_{x}u^3+825v_{x}v^2u_{x}u+57750v_{x}vu_{xxx}u+21v^6
\\ \nonumber
\phantom{H_3  =}{}
+198v^5u+660v^4u^2-7425v^3u_{xx}u+440v^3u^3+44550v^2u_{4x}u-11550v^2u_{x}^2u
\\ \nonumber
\phantom{H_3  =}{}  +
2640v^2u^4-111375vu_{6x}u-4950vu_{xx}^2u-59400vu_{x}^2u^2+3168vu^5
\big).
\end{gather*}

Taking into the account a simple form of the f\/irst Hamiltonian it is
possible to guess the
f\/irst  Hamiltonian structure
\begin{gather*}
\frac{d }{d t} \left(\begin{array}{c}
u \\ v
\end{array} \right) = P  \left ( \begin{array}{c}
\dfrac{\delta H_1}{\delta u} \vspace{2mm} \\
  \dfrac{\delta H_1}{\delta v}
\end{array}\right ) =
\left(\begin{array}{cc}
 3 \partial^{3} +\partial u + u \partial   &   0   \\
  0   &   3\partial^{3} - 2(\partial v+ v \partial)
\end{array}
\right) \left ( \begin{array}{c}
 \dfrac{\delta H_1}{\delta u}  \vspace{2mm} \\
 \dfrac{\delta H_1}{\delta v}
\end{array}\right ).
\end{gather*}

In order to def\/ine  the second Hamiltonian structure we f\/irst found
the recursion operator.
We   used  the technique described in \cite{gora1} and we found the
following tenth-order recursion operator
\begin{gather*}
R  =
\left(\begin{array}{cc}
 - \frac{18}{125}\partial^{10} + \mbox{268 terms} &
-\frac{11}{375}\partial^{10} + \mbox{268 terms} \vspace{2mm} \\
  -\frac{11}{365}\partial^{10} + \mbox{268 terms}   &
-\frac{7}{375}\partial^{10} + \mbox{268 terms}
\end{array}\right).
\end{gather*}

Next we assumed that this operator could be  factorized as $R = J^{-1}
P $ where $J^{-1}$ is the inverse
Hamiltonian  operator.
Due to the diagonal form of the f\/irst Hamiltonian structure it is easy
to carry out such procedure and
as a result we obtained  the second Hamiltonian structure
\begin{gather*}
J^{-1} \frac{d }{d t} \left(\begin{array}{c}
u \\ v \end{array} \right ) =
 \left ( \begin{array}{c}
\frac{\delta H_4}{\delta u} \\
 \frac{\delta H_4}{\delta v}
\end{array}\right ),
\end{gather*}
where
\begin{gather*}
 H_4 = \int dx \big ( 21u_{10x}u + 26v_{10x}u  + \mbox{95 terms})
\end{gather*}
and the explicit form of $H_4$ and  $J^{-1}$ is given in the appendix.

\section{The derivation of the Lax representation}

The Lax operator of equations~\eqref{eq18} has been discovered
accidentally during the investigations of
the supersymmetric Manin--Radul hierarchy.
This hierarchy  can be described by the supersymmetric Lax operator
\begin{gather}\label{eq24}
 L= {\cal D} + f_{0} + \sum_{j=1}^{\infty} b_j \partial^{-j} {\cal D}
+ \sum_{j=1}^{\infty} f_j\partial^{-j},
\end{gather}
where the coef\/f\/icients $b_j$, $f_j$ are bosonic and fermionic
superf\/ield functions, respectively.
We shall use the following notation throughout the paper: $\partial$
and ${\cal D}= \frac{\partial}{
\partial \theta} + \theta \partial$. As usual, $(x,\theta)$ denotes
$N=1$ superspace coordinates. For any super pseudodif\/ferential
operator ${\cal A} = \sum\limits_{j} a_{j/2} {\cal D}^{j}$ the
subscripts  $\pm$ denote its purely
dif\/ferential part $ {\cal A}_{+}= \sum\limits_{j \geq 0} a_{j/2} {\cal
D}^j$ or its purely pseudo-dif\/ferential part
$ {\cal A}_{-}= \sum\limits_{j \geq 1} a_{-j/2} {\cal D}^{-j}$  respectively.
 For any ${\cal A}$ the super-residuum is def\/ined as ${\rm Res}\,
{\cal A} = a_{-1/2}$.

The constrained $(r,m)$ supersymmetric Manin--Radul hierarchy
\cite{Aratyn1} is def\/ined by the fol\-lowing Lax operator
\begin{gather*}
 L= {\cal D}^{r}   + \sum_{j=0}^{r-1} \Psi_{j/2} {\cal D}^{j}  +
\sum_{j=0}^{m}\Upsilon_{\frac{m-j}{2}}{\cal D}^{-1} \Psi_{j/2}.
\end{gather*}
This hierarchy for even $r$ has been widely studied in the literature
in contrast to the odd $r$ which is
less known. Further we will consider this hierarchy for odd $r=3,5$
and  $m=0$, $\Upsilon=\Psi=0$.

The Lax operator for  $r=3$ and  $m=0$, $\Upsilon=\Psi=0$ has been
considered recently by Tian and Liu \cite{Liu}
\begin{gather*}
 L = {\cal D}^3 + \Phi,
\end{gather*}
where $\Phi$ is a superfermion function. Let us  consider  the
following tower of equations
\begin{gather*}
 L_{t,k}= 9 \big[L , L^{k/3}_{+} \big].
\end{gather*}
The f\/irst four  consistent nontrivial equations are
\begin{gather*}
 \Phi_{t,2} =\Phi_{x},
\\
  \Phi_{\tau,7} = \big( \Phi_{1,xx} + \tfrac{1}{2}\Phi_{1}^2
+3\Phi\Phi_{x} \big)_{x},
\\
 \Phi_{t,10}=\Phi_{5x} + 5 \Phi_{xxx}\Phi_{1} +
5\Phi_{xx}\Phi_{1,x} + 5 \Phi_{x}\Phi_{1}^{2},
\\
 \Phi_{\tau,11}  =  \Phi_{1,5x} + 3\Phi_{1,xxx}\Phi_{1} +
6\Phi_{1,xx}\Phi_{1,x} +
2\Phi_{1,x} \Phi_{1}^{2} \nonumber \\
  \phantom{\Phi_{\tau,11}  = }{}  -3\Phi_{4x}\Phi  -
2\Phi_{xxx}\Phi_{x} -6\Phi_{xx}\Phi\Phi_{1}
-6\Phi_{x}\Phi\Phi_{1,x},
\end{gather*}
where $t$ is a usual time while $\tau$ is an odd time.

The third equation in the hierarchy in the component $\Phi = \xi +
\theta w$ reads
\begin{gather*} 
 \xi_{t}  =  \xi_{5x} + 5w\xi_{xxx} + 5w_{x}\xi_{xx} + 5u^{2}\xi_{x}, \\
 w_{t}  =  w_{5x} + 5ww_{xxx} + 5w_{x}w_{xx} +5w^{2}w_{x} - 5\xi_{xxx}\xi_{x}
\end{gather*}
and it is a supersymmetric generalization of the Sawada--Kotera
equation. This equation is a~bi-Hamiltonian system with odd
supersymmetric Poisson brackets~\cite{Pop2}.
The proper Hamiltonian operator which satisf\/ies the Jacobi identity
and generates the supersymmetric $N=1$ Sawada--Kotera equation  is
\begin{gather*}
 \Phi_{t,10}  = P\frac{\delta H_1}{\delta \Phi},
\end{gather*}
where $H_1=\int \Phi \Phi_{x} \, d x  d\theta$ and
\begin{gather*}
 P = \big( {\cal D} \partial^{2} + 2 \partial  \Phi  +  2\Phi \partial
 + {\cal D} \Phi  {\cal D}\big)
 \partial^{-1}   \big( {\cal D} \partial^{2} + 2 \partial  \Phi  +
2\Phi \partial  + {\cal D} \Phi
 {\cal D} \big).
\end{gather*}
The implectic operator for this equation  was def\/ined in \cite{Pop2}   as
\begin{gather}
J\Phi_t  =  \frac{\delta H_3}{\delta \Phi}, \qquad
J  =   \partial_{xx} +({\cal D}\Phi) - \partial^{-1} ({\cal
D}\Phi)_{x} + \partial^{-1}\Phi_{x} {\cal D} +
\Phi_{x}\partial^{-1}{\cal D},\label{eq36}
\end{gather}
where
\begin{gather*}
H_3 = \int  dx d\theta \big( \Phi_{7x}\Phi + 8\Phi_{xxx}\Phi({\cal
D}\Phi)_{xx} +
\Phi_{x}\Phi (4({\cal D}\Phi)_{4x}\nonumber\\
 \phantom{H_3 =}{} + 20({\cal D}\Phi)_{xx}({\cal D}\Phi) + 10 ({\cal
D}\Phi)_{x}^{2} + \tfrac{8}{3}({\cal D} \Phi)^{3}) \big).
 \end{gather*}
This supersymmetric equation possesses an inf\/inite number  of
conserved charges~\cite{Liu} which are generated by the supertrace
formula of the Lax operator. However, these charges are not reduced to
the known conserved
charges in the bosonic limit. Hence we can not in general conclude
that from the supersymmetric integrability
follow the integrability of the bosonic sector.

Let us now consider the Lax operator \eqref{eq24} for $r=5$ and $m=0$
\begin{gather*}
 L=  {\cal D}^{5} + \tfrac{1}{3} (\partial U + U \partial ) -
\tfrac{1}{3}{\cal D}  V {\cal D},
\end{gather*}
where $U$ and $V$ are superfermionic functions $U = \xi + \theta u$,
$V =\psi + \theta v$.
The f\/irst nontrivial equations in the hierarchy generated by this Lax
operator is given as
\begin{gather*}
 L_t= \big[ L^{6/5}_{+},L \big],
\\
 \nonumber
 U_t = 4 U_{xxx} + 3V_{xxx} - 2U_{x}({\cal D}U + {\cal D}V)  +
U(6{\cal D}U_{x} + 2{\cal D}V_{x})
- V_{x}{\cal D}V + V(3{\cal D}U_{x} + {\cal D}V_{x}), \\ \nonumber
 V_t = 3U_{xxx} +V_{xxx} +8U_{x}{\cal D}U - U(8{\cal D}U_{x} - 6{\cal
D}V_{x}) +V_{x}(4{\cal D}U + {\cal D}V) -
V(4{\cal D}U_{x} + 3{\cal D}V_{x}).
\end{gather*}
The bosonic sector of the latter system where $\xi=0$, $\psi=0$ gives
us the system of two interacted KdV type equations discovered by
Drinfel'd--Sokolov.

Interestingly, the Lax operator equations~\eqref{eq36} did not reduce
in the bosonic sector to our Lax operator \eqref{eq18}, however, its
second power reduces that one can  easy  verify.
As we checked, this system possesses  the  same properties as the
supersymmetric Sawada--Kotera equation.
Namely, this model, due to the Lax representation, has an inf\/inite
number of conserved quantities, which are not reduced to the usual
conserved charges in the bosonic limit. For example, the f\/irst two
conserved charges are
\begin{gather}
 H_1  =  \int dx d{\theta}  (4U_{x}U + 6V_{x}V + V_{x}V),\nonumber \\
 H_2  =  \int dx d{\theta}  \big(75U_{xxx}U + 32U_{x}U({\cal D}U) -
24U_{x}U({\cal D}V) +
90V_{xxx}U + 30V_{xxx}V  \nonumber \\
\phantom{H_2  =}{}  +36V_{x}U({\cal D}U) - 6V_{x}U({\cal D}V) -
4V_{x}V({\cal D}V) - 30VU({\cal D}V_{x})\big).\label{eq40}
\end{gather}
We  found the following odd Hamiltonian structure for our
supersymmetric equation \eqref{eq40}
\begin{gather*}
\frac{d }{d t} \left(\begin{array}{c}
 U  \\ V
\end{array} \right) =
\left(\begin{array}{cc}
  \frac{1}{30}    &   \frac{1}{10}  \vspace{2mm} \\
  -\frac{1}{10}   &    -\frac{2}{15}
\end{array}
\right) \left ( \begin{array}{c}
  \dfrac{\delta H_2}{\delta u}  \vspace{2mm} \\
  \dfrac{\delta H_2}{\delta v}
\end{array}\right ).
\end{gather*}
Unfortunately, we have been not able to found second Hamiltonian
structure for our superequation.

\appendix

\section{Appendix}

The conserved quantity $H_4$ is
\begin{gather*}\nonumber
H_4  =  u_{10x}u
-\tfrac{7312}{315}u_{5x}u_{xxx}u
-\tfrac{6638}{315}u_{4x}^{2}u-\tfrac{2032}{945}u_{xxx}^{2}u^{2}
 -\tfrac{3496}{315}u_{xxx}u_{xx}u_{x}u -\tfrac{584}{945}u_{xx}^{3}u
\\ \nonumber
\phantom{H_4=}{}   +\tfrac{2416}{2025}u_{xx}^{2}u^{3}
-\tfrac{6196}{6075}u_{x}^{4}u-\tfrac{448}{1215}u_{x}^{2}u^{4}+\tfrac{8704}{3189375}u^{7}+\tfrac{26}{21}v_{10x}u
+\tfrac{8}{21}v_{10x}v
-\tfrac{338}{315}v_{8x}vu \\ \nonumber
\phantom{H_4=}{} -\tfrac{1612}{315}v_{7x}v_{x}u
-\tfrac{832}{63}v_{6x}v_{xx}u+\tfrac{52}{105}v_{6x}v^{2}u
-\tfrac{832}{35}v_{5x}v_{xxx}u
+\tfrac{1234}{315}v_{5x}v_{xxx}v+\tfrac{208}{63}v_{5x}v_{x}vu\\
\nonumber
\phantom{H_4=}{}
+\tfrac{416}{1575}v_{5x}vu_{x}u-\tfrac{494}{35}v_{4x}^{2}u+\tfrac{1121}{315}v_{4x}^{2}v
+\tfrac{10127}{315}v_{4x}u_{4x}u + \tfrac{1976}{315}v_{4x}v_{xx}vu
+\tfrac{572}{105}v_{4x}v_{x}^{2}u\\ \nonumber
\phantom{H_4=}{}
+\tfrac{6136}{2835}v_{4x}v_{x}u_{x}u - \tfrac{572}{4725}v_{4x}v^{3}u
-\tfrac{416}{14175}v_{4x}v^{2}u^{2}-\tfrac{416}{945}v_{xxx}^{2}u^{2}-\tfrac{53}{315}v_{xxx}^{2}v^{2}
+\tfrac{754}{189}v_{xxx}^{2}vu  \\ \nonumber
\phantom{H_4=}{}
+\tfrac{23582}{315}v_{xxx}u_{5x}u\!+\tfrac{416}{189}v_{xxx}v_{xx}u_{x}u\!
+ \tfrac{1144}{63}v_{xxx}v_{xx}v_{x}u\!-\tfrac{872}{945}v_{xxx}v_{xx}v_{x}v\!
-\tfrac{3952}{4725}v_{xxx}v_{x}v^{2}u \\ \nonumber
\phantom{H_4=}{}
 -\tfrac{130}{567}v_{xxx}v_{x}vu^{2}-\tfrac{52}{945}v_{xxx}vu_{xxx}u
+\tfrac{104}{27}v_{xx}^{3}u-\tfrac{74}{945}v_{xx}^{3}v
-\tfrac{416}{945}v_{xx}^{2}u_{xx}u +
\tfrac{2288}{14175}v_{xx}^{2}u^{3} \\ \nonumber
\phantom{H_4=}{}
-\tfrac{88}{2025}v_{xx}^{2}v^{3} - \tfrac{7592}{14175}v_{xx}^{2}v^{2}u
-\tfrac{26}{135}v_{xx}^{2}vu^{2} +
\tfrac{2782}{45}v_{xx}u_{6x}u+\tfrac{2548}{135}v_{xx}u_{xx}^{2}u
-\tfrac{3484}{2025}v_{xx}v_{x}^{2}vu
\\ \nonumber
\phantom{H_4=}{}
-\tfrac{2236}{2835}v_{xx}v_{x}u_{xxx}u
+\tfrac{104}{6075}v_{xx}v^{4}u +\tfrac{208}{42525}v_{xx}v^{3}u^{2}
+\tfrac{104}{405}v_{xx}v^{2}u_{xx}u -\tfrac{494}{2025}v_{x}^{4}u
\\ \nonumber
\phantom{H_4=}{}
 +\tfrac{233}{6075}v_{x}^{4}v +\tfrac{26}{567}v_{x}^{3}u_{x}u
-\tfrac{1976}{4725}v_{x}^{2}u_{xx}u^{2}
-\tfrac{1924}{1575}v_{x}^{2}u_{x}^{2}u
-\tfrac{832}{14175}v_{x}^{2}u^{4}-\tfrac{38}{6075}v_{x}^{2}v^{4}
\\ \nonumber
\phantom{H_4=}{}
+\tfrac{52}{1215}v_{x}^{2}v^{3}u
-\tfrac{208}{8505}v_{x}^{2}v^{2}u^{2}+\tfrac{286}{567}v_{x}^{2}vu_{xx}u
-\tfrac{3328}{42525}v_{x}^{2}vu^{3} +\tfrac{728}{45}v_{x}u_{7x}u
-\tfrac{8164}{315}v_{x}u_{4x}u_{x}u
  \\ \nonumber
\phantom{H_4=}{}
-\tfrac{10348}{945}v_{x}u_{xxx}u_{xx}u
-\tfrac{11518}{4725}v_{x}u_{x}^{3}u +\tfrac{3952}{4725}v_{x}vu_{5x}u
-\tfrac{12688}{14175}v_{x}vu_{xx}u_{x}u -\tfrac{68}{3189375}v^{7}
  \\ \nonumber
\phantom{H_4=}{}
-\tfrac{104}{455625}v^{6}u-\tfrac{416}{455625}v^{5}u^{2}
+\tfrac{104}{8505}v^{4}u_{xx}u
-\tfrac{832}{637875}v^{4}u^{3}-\tfrac{1144}{14175}v^{3}u_{4x}u
+\tfrac{1664}{42525}v^{3}u_{xx}u^{2}  \\ \nonumber
\phantom{H_4=}{}
+\tfrac{416}{14175}v^{3}u_{x}^{2}u
+\tfrac{1664}{637875}v^{3}u^{4}
+\tfrac{338}{945}v^{2}u_{6x}u+\tfrac{208}{4725}v^{2}u_{4x}u^{2}
+\tfrac{104}{4725}v^{2}u_{xxx}u_{x}u
  \\ \nonumber
\phantom{H_4=}{}
-\tfrac{104}{4725}v^{2}u_{xx}^{2}u +\tfrac{1664}{14175}v^{2}u_{xx}u^{3}
+\tfrac{10816}{42525}v^{2}u_{x}^{2}u^{2}
+\tfrac{3328}{455625}v^{2}u^{5} -\tfrac{247}{315}vu_{8x}u
+\tfrac{208}{105}vu_{6x}u^{2}  \\ \nonumber
\phantom{H_4=}{}
-\tfrac{10868}{945}vu_{5x}u_{x}u -\tfrac{14872}{315}vu_{4x}u_{xx}u
+\tfrac{4576}{4725}vu_{4x}u^{3}-\tfrac{29692}{945}vu_{xxx}^{2}u
+\tfrac{102128}{14175}vu_{xxx}u_{x}u^{2} \\
\phantom{H_4=}{}
 +\tfrac{65416}{14175}vu_{xx}^{2}u^{2}+\tfrac{123682}{14175}vu_{xx}u_{x}^{2}u
+\tfrac{1664}{6075}vu_{xx}u^{4}
+\tfrac{4576}{6075}vu_{x}^{2}u^{3}
+\tfrac{3328}{455625}vu^{6}.
\end{gather*}

The inverse Hamiltonian operator $ J^{-1} $ has the following form
\begin{gather*}
J^{-1} =
\left(\begin{array}{cc}
  J_{1,1}^{-1}   &    J_{1,2}^{-1}  \vspace{1mm} \\
  - \big(J_{1,2}^{-1}\big)^{*}   &   J_{2,2}^{-1}
\end{array}\right ),
\end{gather*}
where
\begin{gather*}\nonumber
 J_{1,1}^{-1}  = -\tfrac{3}{125}\partial^7 + a_{1,1,5}\partial^5 +
a_{1,1,3}\partial^3 +
a_{1,1,1}\partial + b_{1,1}\partial^{-1} +
      b_{1,1,1}\partial^{-1} b_{1,1,2} - \mbox{h.c.},\\ \nonumber
J_{1,2}^{-1}  =  -\tfrac{11}{375}\partial^7 + \sum_{i=0}^5
a_{1,2,i}\partial^{i}  + b_{1,2}\partial^{-1} +
\partial^{-1}c_{1,2} + b_{1,2,1}\partial^{-1}b_{1,2,2}, \\
J_{2,2}^{-1}   =  -\tfrac{7}{750}\partial^7 + a_{2,2,5}\partial^5 +
a_{2,2,3}\partial^3 +
a_{2,2,1}\partial + b_{2,2}\partial^{-1} +
      b_{2,2,1}\partial^{-1} b_{2,2,2} - \mbox{h.c.},
\\
\nonumber
a_{1,1,5}  =  (-43u+14v)/1125, \\ \nonumber
a_{1,1,3}  =  (225u_{xx}-448u^2-690v_{xx}-77v^2-42vu)/16875, \\ \nonumber
a_{1,1,1}  =   \big(-3150u_{4x}-1800u_{xx}u+2397u_{x}^2-888u^3+720v_{4x}-2520v_{xx}u
 +420v_{xx}v \\ \nonumber
  \phantom{a_{1,1,1}  =}{}
   +798v_{x}^2-1512v_{x}u_{x}+56v^3-126v^2u-728vu^2\big)/101250,   \\ \nonumber
b_{1,1}  =  \big(-6750u_{6x}-9540u_{4x}u-19080u_{xxx}u_{x}-14310u_{xx}^2-5520u_{xx}u^2
-5520u_{x}^2u-352u^4 \\ \nonumber
 \phantom{b_{1,1}  =}{}
   -4050v_{6x}-3780v_{4x}u+1890v_{4x}v-810v_{3x}u_{x}
+6480v_{3x}v_{x}+3510v_{xx}^2+2160v_{xx}u_{xx} \\ \nonumber
\phantom{b_{1,1}  =}{}
   -2160v_{xx}u^2-540v_{xx}v^2\!
-1440v_{xx}vu\!+720v_{x}^2u-810v_{x}^2v\!+5940v_{x}u_{xxx}\!-1080v_{x}u_{x}u
\\ \nonumber
\phantom{b_{1,1}  =}{}
   -2160v_{x}vu_{x}+18v^4+96v^3u-1080v^2u_{xx}-288v^2u^2
 +2970vu_{4x}-1080vu_{xx}u \\ \nonumber
\phantom{b_{1,1}  =}{}
   -540vu_{x}^2-576vu^3\big)/759375, \\ \nonumber
b_{1,1,1}\partial^{-1}b_{1,1,2}  =  \big(- 990u_{4x} - 1020u_{xx}u -
510u_{x}^2-128u^3-630v_{4x} - 420v_{xx}u \\
\nonumber
\phantom{b_{1,1,1}\partial^{-1}b_{1,1,2}  =}{}
 +210v_{xx}v+210v_{x}^2+210v_{x}u_{x} - 14v^3-56v^2u+210vu_{xx}\\
\phantom{b_{1,1,1}\partial^{-1}b_{1,1,2}  =}{}
 -168vu^2\big) \partial^{-1} (4u + 3v)/759375,
\\
\nonumber
a_{1,2,5}  =  (-38u+19v)/1125, \\ \nonumber
a_{1,2,4}  =  (-36u+68v)_{x}/1125, \\ \nonumber
a_{1,2,3}  =  \big({-}45u_{xx}-124u^2+515v_{xx}-31v^2-36vu\big)/5625,
\\ \nonumber
a_{1,2,2}  =  \big(885u_{3x}-656u_{x}u+1680v_{3x}+178v_{x}u-354v_{x}v-332vu_{x})/16875,
\\ \nonumber
a_{1,2,1}  =  \big(2520u_{4x}-1428u_{xx}u-1599u_{x}^2-328u^3+2610v_{4x}-156v_{xx}u
\\ \nonumber
\phantom{a_{1,2,1}  =}{}
 -957v_{xx}v-876v_{x}^2-1716v_{x}u_{x}+41v^3+114v^2u-1371vu_{xx}-228vu^2\big)/50625,
\\ \nonumber
a_{1,2,0}  =  \big(1575u_{5x}+534u_{3x}u+276u_{xx}u_{x}-128u_{x}u^2+1080v_{5x}+198v_{3x}u
\\ \nonumber
\phantom{a_{1,2,0}  =}{}
   -534v_{3x}v-873v_{xx}u_{x}-1206v_{xx}v_{x}-1338v_{x}u_{xx}+84v_{x}u^2 \\
\phantom{a_{1,2,0}  =}{}
   +96v_{x}v^2+166v_{x}vu+128v^2u_{x}-792vu_{3x}+8vu_{x}u\big)/50625,
\\
\nonumber
b_{1,2}  =   \big(3375u_{6x}+4770u_{4x}u+9540u_{3x}u_{x}+7155u_{xx}^2+2760u_{xx}u^2
\\ \nonumber
\phantom{b_{1,2}  =}{}
   +2760u_{x}^2u+176u^4+2025v_{6x}+1890v_{4x}u-945v_{4x}v+405v_{3x}u_{x}-3240v_{3x}v_{x}
\\ \nonumber
\phantom{b_{1,2}  =}{}
   -1755v_{xx}^2-1080v_{xx}u_{xx}+1080v_{xx}u^2+270v_{xx}v^2+720v_{xx}vu-360v_{x}^2u
\\ \nonumber
\phantom{b_{1,2}  =}{}
   +405v_{x}^2v-2970v_{x}u_{3x}+540v_{x}u_{x}u+1080v_{x}vu_{x}-9v^4-48v^{3}u \\
\phantom{b_{1,2}  =}{}
   +540v^2u_{xx}+144v^2u^2-1485vu_{4x}+540vu_{xx}u+270vu_{x}^2+288vu^3\big)/759375,
   \\ \nonumber
c_{1,2}  =  \big({-}4050u_{6x}-3780u_{4x}u-14310u_{3x}u_{x}-9045u_{xx}^2-2160u_{xx}u^2
 -3780u_{x}^2u -144u^4 \\ \nonumber
\phantom{c_{1,2}  =}{}
   -2700v_{6x}\!-2160v_{4x}u+1530v_{4x}v\!-4320v_{3x}u_{x}\!
+3060v_{3x}v_{x}+2295v_{xx}^2\!  -1080v_{xx}u_{xx} \\ \nonumber
\phantom{c_{1,2}  =}{}
  -2160v_{xx}u^2-390v_{xx}v^2
-540v_{xx}vu-270v_{x}^2u-390v_{x}^2v+1080v_{x}u_{3x}-4320v_{x}u_{x}u
\\ \nonumber
\phantom{c_{1,2}  =}{}
   -540v_{x}vu_{x}+11v^4+72v^3u-540v^2u_{xx}+144v^2u^2+1890vu_{4x}-1440vu_{xx}u
\\ \nonumber
\phantom{c_{1,2}  =}{}
  -360vu_{x}^2-192vu^3\big)/759375, \\ \nonumber
b_{1,2,1}\partial^{-1}b_{1,2,2}  =  \big( \big({-} 990u_{4x}
-1020u_{xx}u-510u_{x}^2-128u^3-630v_{4x}
 -420v_{xx}u+210v_{xx}v \\ \nonumber
 \phantom{b_{1,2,1}\partial^{-1}b_{1,2,2}  =}{}
  +210v_{x}^2+210v_{x}u_{x}+14v^3-56v^2u+210vu_{xx}-168vu^2\big)\partial^{-1}(3u+v)\\
\nonumber
  \phantom{b_{1,2,1}\partial^{-1}b_{1,2,2}  =}{}
   -(4u+3v)\partial^{-1}\big(630u_{4x}+420u_{xx}u+525u_{x}^2+56u^3+360v_{4x}-150v_{xx}v
\\
  \phantom{b_{1,2,1}\partial^{-1}b_{1,2,2}  =}{}
  -75v_{x}^2+8v^3+42v^2u-210vu_{xx}+56vu^2\big)\big)/759375,
   \\
\nonumber
a_{2,2,5}  =  (-9u+7v)/1125, \\ \nonumber
a_{2,2,3}  =  \big(540u_{xx}-122u^2-30v_{xx}-33v^2-28vu\big)/16875, \\ \nonumber
a_{2,2,1}  =  \big({-}1170u{4x}+900u_{xx}u+1203u_{x}^2-120u^3+540v_{4x}-150v_{xx}v
\\ \nonumber
\phantom{a_{2,2,1}  =}{}
   +162v_{x}^2-108v_{x}u_{x}+30v^3+110v^2u-570vu_{xx}+120vu^2\big)/101250,
\\ \nonumber
b_{2,2}  =  \big(4050u_{6x}+3780u_{4x}u+14310u_{3x}u_{x}+9045u_{xx}^2+2160u_{xx}u^2+3780u_{x}^2u
\\ \nonumber
\phantom{b_{2,2}  =}{}
   +144u^4+2700v_{6x}+2160v_{4x}u-1530v_{4x}v+4320v_{3x}u_{x}-3060v_{3x}v_{x}
\\ \nonumber
\phantom{b_{2,2}  =}{}
   -2295v_{xx}^2+1080v_{xx}u_{xx}+2160v_{xx}u^2+390v_{xx}v^2+540v_{xx}vu+270v_{x}^2u+390v_{x}^2v
\\ \nonumber
\phantom{b_{2,2}  =}{}
   -1080v_{x}u_{3x}+4320v_{x}u_{x}u+540v_{x}vu_{x}-11v^4-72v^3u+540v^2u_{xx}
\\ \nonumber
\phantom{b_{2,2}  =}{}
   -144v^2u^2-1890vu_{4x}+1440vu_{xx}u+360vu_{x}^2+192vu^3\big)/1518750,
\\ \nonumber
b_{2,2,1}\partial^{-1}b_{2,2,2}  =  \big({-}
630u_{4x}-420u_{xx}u-525u_{x}^2-56u^3-360v_{4x} \\
\phantom{b_{2,2,1}\partial^{-1}b_{2,2,2}  =}{}
   +150v_{xx}v+75v_{x}^2-8v^3-42v^2u+210vu_{xx}-56vu^2\big)\partial^{-1}(3u+v)/759375
.  
\end{gather*}

\pdfbookmark[1]{References}{ref}
\LastPageEnding

\end{document}